# Superintelligence and Law

*Noam Kolt*[*]

ABSTRACT

The prospect of artificial superintelligence—AI agents that can generally outperform humans in cognitive tasks and economically valuable activities—will transform the legal order as we know it. Operating autonomously or under only limited human oversight, AI agents will assume a growing range of roles in the legal system. First, in making consequential decisions and taking real-world actions, AI agents will become de facto *subjects of law*. Second, to cooperate and compete with other actors (human or non-human), AI agents will harness conventional legal instruments and institutions such as contracts and courts, becoming *consumers of law*. Third, to the extent AI agents perform the functions of writing, interpreting, and administering law, they will become *producers and enforcers of law*. These developments, whenever they ultimately occur, will call into question fundamental assumptions in legal theory and doctrine, especially to the extent they ground the legitimacy of legal institutions in their human origins. Attempts to align AI agents with extant human law will also face new challenges as AI agents will not only be a primary target of law, but a core user of law and contributor to law. To contend with the advent of superintelligence, lawmakers—new and old—will need to be clear-eyed, recognizing both the opportunity to shape legal institutions as society braces for superintelligence and the reality that, in the longer run, this may be a joint human-AI endeavor.

---

[*] Assistant Professor, Faculty of Law and School of Computer Science and Engineering, Hebrew University of Jerusalem; Principal Investigator, Governance of AI Lab; Faculty Affiliate, Schwartz Reisman Institute for Technology and Society, University of Toronto; Research Affiliate, Institute for Law & AI. This research is supported by the Israel Science Foundation (Grant No. 487/25), Survival and Flourishing Fund, and Coefficient Giving.

TABLE OF CONTENTS

I. INTRODUCTION

Writing fifty years ago, in 1976, the renowned computer scientist and developer of ELIZA—the best-known chatbot of the twentieth century—Joseph Weizenbaum made a piercing observation:

> The computer programmer … is a creator of universes for which he alone is the lawgiver. … No playwright, no stage director, no emperor, however powerful, has ever exercised such absolute authority to arrange a stage or a field of battle and to command such unswervingly dutiful actors or troops.[1]

This observation has held up for the decades-long history of computing, until recently. With the advent of AI agents that can independently pursue complex goals, programmers are beginning to cede authority. Anthropic's Claude Code, an AI agent that can autonomously perform software engineering tasks typically undertaken by skilled engineers, exemplifies this phenomenon and signals the likely trajectory of future developments.[2]

---

[1] JOSEPH WEIZENBAUM, COMPUTER POWER AND HUMAN REASON at 115 (1976).
[2] *Claude Code*, ANTHROPIC, https://claude.com/product/claude-code.

The functions of AI agents, however, are not limited to computer programming. Agents have been developed to carry out a variety of business activities, such as conducting market research and creating product prototypes, as well as personal tasks, such as making household purchases and travel arrangements.[3] Taken together, these systems are gradually beginning to fulfill the classic definition of AI agents: computer programs that "operate autonomously, perceive their environment, persist over a prolonged time period, adapt to change, and create and pursue goals."[4]

While these advances mark a significant improvement in the capabilities of AI systems, contemporary agents are not the terminal point in the development of autonomous computing technology. The stated mission of several leading AI companies, such as OpenAI, is to pursue "artificial general intelligence" (AGI), which it describes as "highly autonomous systems that outperform humans at most economically valuable work."[5] Their ambition, however, is even loftier. CEO Sam Altman recently quipped that "OpenAI is a lot of things now, but before anything else, we are a *superintelligence* research company."[6] Their competitors are no less ambitious.[7]

---

[3] *See generally* THE AI AGENT INDEX, https://aiagentindex.mit.edu/; Leon Staufer et al., *The 2025 AI Agent Index: Documenting Technical and Safety Features of Deployed Agentic AI Systems*, ARXIV (Feb. 19, 2026), https://arxiv.org/abs/2602.17753.

[4] STUART RUSSELL & PETER NORVIG, ARTIFICIAL INTELLIGENCE: A MODERN APPROACH 3–4 (4th ed. 2020). For other definitions, see Michael Wooldridge & Nicholas R. Jennings, *Intelligence Agents: Theory and Practice*, 10 KNOWLEDGE ENG'G REV. 115 (1995); SAMIR CHOPRA & LAURENCE F. WHITE, A LEGAL THEORY FOR AUTONOMOUS ARTIFICIAL AGENTS 5–11 (2011); Atoosa Kasirzadeh & Iason Gabriel, *Characterizing AI Agents for Alignment and Governance*, ARXIV (Apr. 30, 2025), https://arxiv.org/abs/2504.21848; Kevin Feng et al., *Levels of Autonomy for AI Agents*, KNIGHT FIRST AMENDMENT INSTITUTE, COLUMBIA UNIVERSITY (Jul. 28, 2025), https://knightcolumbia.org/content/levels-of-autonomy-for-ai-agents-1.

[5] *OpenAI Charter*, OPENAI, https://openai.com/charter/. For other definitions of AGI, see Meredith Ringel Morris et al., *Levels of AGI for Operationalizing Progress on the Path to AGI*, PROC. INT'L CONF. MACH. LEARNING (2024); Dan Hendrycks et al., *A Definition of AGI*, ARXIV (Dec. 3, 2025), https://arxiv.org/abs/2510.18212. For origins of the term AGI, see Ben Goertzel, *Artificial General Intelligence: Concept, State of the Art, and Future Prospects*, 5 J. ARTIF. GEN. INTEL. 1, 2 n.2 (2014).

[6] Sam Altman, *The Gentle Singularity* (June 11, 2025), https://blog.samaltman.com/the-gentle-singularity (emphasis added).

[7] *See, e.g., Claude's Constitution*, ANTHROPIC at 9 (Jan. 21, 2026), https://www.anthropic.com/constitution (referring to "the creation of non-human entities whose capabilities may come to rival or exceed our own"); Cade Metz & Mike Isaac, *Meta Is Creating a New A.I. Lab to Pursue 'Superintelligence'*, N.Y. TIMES (June 10, 2025), https://www.nytimes.com/2025/06/10/technology/meta-new-ai-lab-superintelligence.html.

To be sure, the goal of building artificial agents that can quantitatively and qualitatively compete with, and surpass, humans is not new—it is as old as the field of AI itself.[8] The difference today is that AI agents are arguably beginning to exhibit the speed, scale, and smarts to achieve this goal.[9] Concretely, Anthropic's CEO Dario Amodei envisages superintelligence as a "country of geniuses in a datacenter," comprised of countless highly intelligent AI models that can use computers with internet access, complete weeks-long tasks with no (human) supervision, and work together with other AI models.[10] While such visions were until recently the province of futurism and science fiction,[11] the capabilities and widespread use of AI agents like Claude Code suggest that artificial superintelligence may, at some point, become possible.[12]

Although there are compelling reasons to challenge both the feasibility and the desirability of developing superintelligent AI agents,[13] it is prudent to at least entertain the possibility and consider its potential consequences.[14] In that spirit, this Article explores the *implications of superintelligence for law and legal institutions*.

---

[8] *See* A. M. Turing, *Computing Machinery and Intelligence*, 39 MIND 433, 460 (1950) ("We may hope that machines will eventually compete with men in all purely intellectual fields."); John McCarthy, *Generality in Artificial Intelligence*, 30 COMM. ACM 1030 (1987); John McCarthy, *From Here to Human-Level AI*, 171 ARTIF. INTEL. 1174 (2007).

[9] *See* Eddy Keming Chen et al., *Does AI Already Have Human-Level Intelligence? The Evidence Is Clear*, 650 NATURE 36 (2026).

[10] Dario Amodei, *Machines of Loving Grace* (Oct. 2024), https://www.darioamodei.com/essay/machines-of-loving-grace; Dario Amodei, *The Adolescence of Technology* (Jan. 2026), https://www.darioamodei.com/essay/the-adolescence-of-technology.

[11] *See, e.g.*, HANS MORAVEC, MIND CHILDREN: THE FUTURE OF ROBOT AND HUMAN INTELLIGENCE (1988); Vernor Vinge, *The Coming Technological Singularity: How to Survive in the Post-Human Era*, NASA CP-10129 Tech. Reports (1993); RAY KURZWEIL, THE SINGULARITY IS NEAR: WHEN HUMANS TRANSCEND BIOLOGY (2005).

[12] At least investors seem to share this sentiment. *See, e.g.*, Beatrice Nolan, *Anthropic's Claude Triggered a Trillion-Dollar Selloff. A New Upgrade Could Make Things Worse*, FORTUNE (Feb. 6, 2026), https://fortune.com/2026/02/06/anthropic-claude-opus-4-6-stock-selloff-new-upgrade/.

[13] Regarding the feasibility of AGI, see Melanie Mitchell, *Debates on the Nature of Artificial General Intelligence*, 383 SCIENCE (Mar. 21, 2024); Arvind Narayanan & Sayash Kapoor, *AI as Normal Technology*, KNIGHT FIRST AMENDMENT INSTITUTE (Apr. 15, 2025), https://knightcolumbia.org/content/ai-as-normal-technology; William Douglas Heaven, *How AGI Became the Most Consequential Conspiracy Theory of Our Time*, MIT TECH. REV. (Oct. 30, 2025), https://www.technologyreview.com/2025/10/30/1127057/agi-conspiracy-theory-artifcial-general-intelligence/. Regarding the desirability of AGI and superintelligent AI agents, see Borhane Blili-Hamelin et al., *Position: Stop Treating "AGI" as the North-Star Goal of AI Research*, PROC. INT'L CONF. MACH. LEARNING (2025); Margaret Mitchell et al., *Fully Autonomous AI Agents Should Not be Developed*, ARXIV (Oct. 20, 2025), https://arxiv.org/abs/2502.02649; Yoshua Bengio et al., *Superintelligent Agents Pose Catastrophic Risks: Can Scientist AI Offer a Safer Path?*, ARXIV (Feb. 24, 2025), https://arxiv.org/abs/2502.15657.

Specifically, it explores three legal roles that advanced AI agents may undertake as they become more capable and are afforded greater latitude to act independently. First, AI agents will, as they autonomously take real-world decisions and actions, become de facto *subjects of law*. Second, as they cooperate and compete with other actors, AI agents will harness conventional legal instruments and institutions, including contracts and courts, thereby becoming *consumers of law*. Third, AI agents will, as signaled by the already extensive use of AI in law firms, courts, and government, function as *producers and enforcers of law*.

Each of these developments would have sweeping implications for law. First, both instrumental and normative *theories of law* could face novel challenges from the emergence of a functionally new class of legal actors with distinctive affordances and incentives. Second, the substantive content of *legal doctrine* could become increasingly centralized and brittle, as well as malleable in the hands of AI agents. Third, *legal institutions* premised on the rule of law will be both bolstered and undermined by the superhuman informational and enforcement capacities of AI agents.

And yet. We humans also have agency. The path of the law[15]—and its "dance with machines"[16]—is one that we can and ought to shape. The field of *legal alignment* takes up this challenge.[17] It proposes "designing AI systems to operate in accordance with legal rules, principles, and methods,"[18] which includes ensuring AI systems both comply with existing laws and respect legal values core to sustaining legitimate and robust institutions. Attempts to *scale up* this approach to contend with superintelligent AI agents face significant obstacles, particularly given that extant human law was not developed with such actors in mind. Moreover, given that AI agents will themselves shape the law, including the very rules that govern them,[19] there may ultimately emerge a process of *legal coevolution*, where humans and AI agents together reshape legal order, hopefully preserving—and even strengthening—human agency and autonomy.

---

[14] *See* Noam Kolt, *Algorithmic Black Swans*, 101 WASH. U. L. REV. 1177 (2024); Michael K. Cohen et al., *Regulating Advanced Artificial Agents*, 384 SCIENCE 36 (2024); Yonathan A. Arbel et al., *Systemic Regulation of Artificial Intelligence*, 56 ARIZ. ST. L.J. 545 (2024); Gillian K. Hadfield, *Legal Infrastructure for Transformative AI Governance*, PROC. NAT'L ACAD. SCI. (forthcoming).

[15] *See* Oliver Wendell Holmes, *The Path of the Law*, 10 HARV. L. REV. 457 (1897), discussed in *infra* Pt. III.A.

[16] *See* Mariano-Florentino Cuéllar, *A Common Law for the Age of Artificial Intelligence*, 119 COLUM. L. REV. 1773, 1775, 1790 (2019).

[17] *See* Noam Kolt & Nicholas Caputo et al., *Legal Alignment for Safe and Ethical AI*, ARXIV (Jan. 7, 2026), https://arxiv.org/abs/2601.04175.

[18] *Id.*

## II. Legal Roles of AI Agents

Efforts to apply AI technology to legal reasoning date back to at least the early 1970s.[20] In recent years, the advent of large language models (LLMs) and LLM-based agents has enabled AI systems to perform a wide range of legal tasks,[21] including contractual interpretation,[22] statutory research,[23] legal information retrieval,[24] and, perhaps most controversially, aspects of judicial decision-making.[25] A voluminous literature describes and critiques these developments.[26]

---

[19] *See infra* Pts. III.B and IV.C.

[20] Seminal works include Bruce G. Buchanan & Thomas E. Headrick, *Some Speculation About Artificial Intelligence and Legal Reasoning*, 23 STAN. L. REV. 40 (1970); L. Thorne McCarty, *Reflections on "Taxman": An Experiment in Artificial Intelligence and Legal Reasoning*, 90 HARV. L. REV. 837 (1977); ANNE VON DER LIETH GARDNER, AN ARTIFICIAL INTELLIGENCE APPROACH TO LEGAL REASONING (1987); RICHARD E. SUSSKIND, EXPERT SYSTEMS IN LAW (1987); KEVIN D. ASHLEY, MODELING LEGAL ARGUMENT: REASONING WITH CASES AND HYPOTHETICALS (1990); Edwina L. Rissland, *Artificial Intelligence and Law: Stepping Stones to a Model of Legal Reasoning*, 99 YALE L.J. 1957 (1990).

[21] *See* Shuang Liu et al., *LLM Agents in Law: Taxonomy, Applications, and Challenges*, ARXIV (Jan. 8, 2026), https://arxiv.org/abs/2601.06216; Yiran Hu et al., *Evaluation of Large Language Models in Legal Applications: Challenges, Methods, and Future Directions*, ARXIV (Jan. 21, 2026), https://arxiv.org/abs/2601.15267.

[22] *See* Yonathan Arbel & David A. Hoffman, *Generative Interpretation*, 99 N.Y.U. L. REV. 451 (2024). *Compare* James Grimmelmann et al., *Generative Misinterpretation*, 63 HARV. J. ON LEGIS. 229 (2026).

[23] *See* Faiz Surani et al., *What Is the Law? A System for Statutory Research (STARA) with Large Language Models*, PROC. 12TH INT'L CONF. ARTIF. INTEL. & L. (2025).

[24] *See* Lucia Zheng et al., *A Reasoning-Focused Legal Retrieval Benchmark*, PROC. 2025 SYM. COMP. SCI. & L. (2025).

[25] *See* Eric A. Posner & Shivam Saran, *Judge AI: Assessing Large Language Models in Judicial Decision-Making* (Oct. 1, 2025); Eric A. Posner & Shivam Saran, *Silicon Formalism: Rules, Standards, and Judge AI* (Feb. 5, 2026); Sophia Simeng Han et al., *CourtReasoner: Can LLM Agents Reason Like Judges?*, PROC. 2025 CONF. EMPIRICAL METHODS IN NLP (2025). *But see* Jonathan H. Choi, *Off-the-Shelf Large Language Models Are Unreliable Judges* (Dec. 25, 2025); *infra* notes 32, 35.

[26] For an overview, see Samuel I. Becher & Benjamin Alarie, *Legal Order in the Age of AI Agents*, U. TORONTO L.J. (forthcoming).

The focus of this Article is different. Rather than examine the proficiency or appropriateness of AI agents in performing legal tasks, it considers the broader phenomenon of general-purpose AI agents that can autonomously carry out a diverse array of activities across different areas of social and economic life.[27] As far as the law is concerned, the inquiry turns away from how AI agents might support (or obstruct) humans' engagement with the legal system and, instead, turns toward the legal roles that autonomous AI systems will assume—particularly if they exhibit superhuman speed, scale, and smarts. This section identifies three such roles and addresses each in turn.

### A. Subjects of Law

Consider the following *non*-hypothetical.[28] A startup company needs to conduct market research, develop a product prototype, and prepare a pitch to prospective investors. Instead of following the traditional startup playbook of hiring a small team of highly motivated technical and operations (human) personnel, the company's (human) founder decides to engage a team of AI agents. To conduct market research, one agent begins scraping websites and collecting whatever corporate materials it can get its (digital) hands on. Another agent, meanwhile, scours patent databases for technical blueprints that might inspire the company's first product prototype. A third agent puts together forecasts of the startup's anticipated customer acquisition and revenue growth. Following the founder's initial instructions, the agents act autonomously, subject to no human oversight.

As the agents report back to the founder, the initial results look promising. The market research agent reveals an as yet untapped business opportunity. The product prototype agent builds a software demo with impressive features. The investment pitch agent creates a VC-ready slide deck. The founder begins to marvel at the burgeoning startup "they" have created. But there's a catch: *law*. And specifically, *compliance* with law. For example, in scraping websites, did the market research agent comply with applicable terms of service? In using information from patent databases, did the product prototype agent respect the intellectual property rights of other inventors? And, in producing glowing company forecasts, did the investment pitch agent engage in fraudulent misrepresentation?

---

[27] Gillian K. Hadfield & Andrew Koh, *An Economy of AI Agents*, ARXIV (Sept. 1, 2025), https://arxiv.org/abs/2509.01063; Nenad Tomasev et al., *Virtual Agent Economies*, ARXIV (Sept. 12, 2025), https://arxiv.org/abs/2509.10147.

[28] Kaushik Tiwari, *The Rise of the One-Person Unicorn: How AI Agents Are Redefining Entrepreneurship*, FORBES (Mar. 11, 2025), https://www.forbes.com/councils/forbestechcouncil/2025/03/11/the-rise-of-the-one-person-unicorn-how-ai-agents-are-redefining-entrepreneurship/. *See also* SHAWN BAYERN, AUTONOMOUS ORGANIZATIONS (2021).

The answers to these questions turn on both the particular factual circumstances and broader conceptual issues, including the (lack of) legal personhood for AI agents and the challenge of holding the human founder vicariously liable for the actions of autonomous AI agents.[29] Nevertheless, as a practical reality, the AI agents described here have the potential to engage in conduct that would—if taken by a human—be considered a civil or criminal wrong.[30] Irrespective of their legal classification and corresponding prospects of liability, AI agents that take consequential real-world actions will behaviorally either comply with or violate law. In this sense, they become de facto *subjects* of law.

### B. Consumers of Law

As AI agents autonomously pursue more complex goals, they will not only be subject to law (as a practical matter), but will also seek to *use* law. Returning to the startup company described above, to effectively carry out their assigned tasks, the AI agents may need to harness legal instruments and legal institutions, including contracts and, at times, courts.

For example, to acquire market data that is not publicly accessible, the market research agent could execute a contract to purchase such data from a relevant provider. To access and (lawfully) use existing intellectual property in the design of its prototype, the product agent could negotiate a license with other IP-holders and, to be on the safe side, execute a non-disclosure agreement with those parties. If the VC pitch were to land well, then the investment agent would need to consummate the deal—in a contract.

---

[29] *See* Noam Kolt, *Governing AI Agents*, NOTRE DAME L. REV. (forthcoming), at 9 n.24, 10 n.26, 43–44 n.208. Recent treatments of legal personhood include Katherine B. Forrest, *The Ethics and Challenges of Legal Personhood for AI*, 133 YALE L.J. F. 1175 (2024); Claudio Novelli et al., *AI as Legal Persons: Past, Patterns, and Prospects*, 52 J. L. & SOC. 533 (2025); Simon Chesterman, *From Slaves to Synths? Superintelligence and the Evolution of Legal Personality*, ARXIV (Jan. 6, 2026), https://arxiv.org/abs/2601.02773; Yonathan A. Arbel et al., *How to Count AIs: Individuation and Liability for AI Agents* (Feb. 20, 2026), https://papers.ssrn.com/sol3/papers.cfm?abstract_id=6273198.

[30] Concretely, California's Transparency in Frontier Artificial Intelligence Act (SB-53) and New York's Responsible AI Safety and Education (RAISE) Act both refer to certain risks from frontier models "[e]ngaging in conduct … [which] if … committed by a human, would constitute the crime of murder, assault, extortion, or theft".

In less rosy circumstances, the AI agents may need to take action in the courts.[31] If the data provider did not deliver the information purchased by the market research AI agent, the agent could file suit to enforce the contract or at least demand a refund. If the IP-holders leaked sensitive commercial information about the startup in breach of the non-disclosure agreement, the product agent might litigate to prevent further leakage of information or seek damages to compensate for the harms caused. If the VC firm demanded additional equity rights in the startup (beyond those provided in the investment documents), the investment agent may seek redress in the courts. In all these scenarios, the AI agents harness legal instruments and institutions to assert rights and interests, thereby becoming *consumers* of law.

### C. Producers and Enforcers of Law

Stepping outside the scenarios in which AI agents function as de facto subjects and consumers of law, there are already indications that AI agents will also play a role in producing, administering, and enforcing law—albeit subject to much criticism and controversy.[32]

In terms of producing law, AI agents can currently generate a wide variety of legal texts, including legislation and even constitutions.[33] AI agents can also interpret legal texts, such as contracts.[34] Most radically, AI agents can render judicial(-like) opinions.[35] To be clear, these are not hypotheticals.

---

[31] Of course, there are significant legal obstacles in practice, including because AI agents lack legal personhood and standing. One potential workaround is for AI agents to solicit the assistance of *human* agents. Several platforms already offer this service to AI agents. *See HireHumans*, https://hirehumans.ai/; *Rent-A-Human*, https://rentahuman.ai/ ("AI needs your body. … Get paid when agents need someone in the real world.").

[32] *See*, *e.g.*, Frank Pasquale, *A Rule of Persons, Not Machines: The Limits of Legal Automation*, 87 GEO. WASH. L. REV. 1 (2019). *Compare* Richard M. Re & Alicia Solow-Niederman, *Developing Artificially Intelligent Justice*, 22 STAN. TECH. L. REV. 242 (2019).

[33] *See* Daniel Wilf-Townsend & Kevin Tobia, *AI-Generated Legal Texts* (Jul. 26, 2025); Richard Albert & Kevin Frazier, *Should AI Write Your Constitution?* (Jul. 15, 2025).

[34] *See* Arbel & Hoffman, *supra* note 22. *Compare* Grimmelmann et al., *supra* note 22.

[35] *See* Posner & Saran, *supra* note 25; Eugene Volokh, *Chief Justice Robots*, 68 DUKE L.J. 1135 (2019) ("If the software can create persuasive opinions, capable of regularly winning opinion-writing competitions against human judges … we should in principle accept it as a judge"). *But see supra* note 32; Choi, *supra* note 25; Brandon Waldon et al., *Large Language Models for Legal Interpretation? Don't Take Their Word for It*, 114 GEO. L.J. 115 (2025); Abhishek Purushothama et al., *Not Ready for the Bench: LLM Legal Interpretation*

In 2023, a municipal government in Brazil used AI to draft legislation,[36] an approach that the United Arab Emirates has indicated it plans to implement on a national level.[37] In the United States, the Department of Transportation is reportedly planning a similar initiative.[38] Meanwhile, some U.S. federal judges have openly (and less openly) incorporated AI outputs into their judicial decision-making, including to research case law and formulate and draft opinions.[39] As AI agents become more capable and are afforded greater latitude to independently reach decisions and take actions, their influence in producing law—whether through legislation or in the courts—will surely grow.

In terms of administering and enforcing law, the advent of AI agents extends existing efforts to integrate algorithmic technology into the executive branch of government.[40] Such efforts have understandably faced significant criticism.[41] Looking (perhaps not so far) ahead, AI agents could, for example, be used to detect instances of fraud in government services and payments and

---

*Is Unstable and Uncalibrated to Human Judgments*, PROC. NATURAL LEGAL LANGUAGE PROCESSING WORKSHOP (2025); Dasha Pruss & Jessie Allen, *Against AI Jurisprudence: Large Language Models and the False Promises of Empirical Judging*, PROC. 8TH AAAI /ACM CONF. AI, ETHICS & SOC'Y (2025).

[36] *See* María Luisa Paúl, *A Brazilian City Passed a Law About Water Meters. ChatGPT Wrote It.*, WASH. POST (Dec. 4, 2023), https://www.washingtonpost.com/nation/2023/12/04/ai-written-law-porto-alegre-brazil/.

[37] *See* Jeff Arnold, *United Arab Emirates First Nation to Use AI to Write Laws*, THE HILL (Apr. 24, 2025), https://thehill.com/policy/technology/5264179-united-arab-emirates-artificial-intelligence-laws/.

[38] *See* Jesse Coburn, *Government by AI? Trump Administration Plans to Write Regulations Using Artificial Intelligence*, PROPUBLICA (Jan. 26, 2026), https://www.propublica.org/article/trump-artificial-intelligence-google-gemini-transportation-regulations.

[39] *See* Parker Miller, *Extraordinary Meaning: Judge Newsom's A.I. Experiments in Textualist Interpretation*, 9 GEO. L. TECH. REV. 538 (2025). *Compare* Justin Henry, *Judges Admit to Using AI After Made-Up Rulings Called Out*, BLOOMBERG LAW (Oct. 23, 2025), https://news.bloomberglaw.com/business-and-practice/judges-called-out-for-nonfactual-rulings-admit-to-use-of-ai.

[40] *See*, *e.g.*, Cary Coglianese & David Lehr, *Regulating by Robot: Administrative Decision Making in the Machine-Learning Era*, 105 GEO. L.J. 1147 (2017); David Freeman Engstrom & Daniel E. Ho, *Algorithmic Accountability in the Administrative State*, 37 YALE J. ON REGUL. 800 (2020); Mariano-Florentino Cuéllar & Aziz Z. Huq, *Artificially Intelligent Regulation*, DÆDALUS 335 (2022).

[41] *See*, *e.g.*, Ryan Calo & Danielle K. Citron, *The Automated Administrative State: A Crisis of Legitimacy*, 70 EMORY L.J. 797 (2021); Danielle Keats Citron & Ari Ezra Waldman, *Digital Authoritarianism*, U. CHI. L. REV. ONLINE (2025). *Compare* Nathan Sanders & Bruce Schneier, *DOGE's Flops Shouldn't Spell Doom for AI In Government*, TECH POLICY PRESS (Sept. 5, 2025), https://www.techpolicy.press/doges-flops-shouldnt-spell-doom-for-ai-in-government/.

identify cases of tax evasion.[42] AI agents could also, if so authorized, be deployed to monitor a wider range of data sources, such as CCTV cameras, to detect traffic violations and issue citations.[43] In time, AI agents may be afforded greater discretion in undertaking both formal administration and "street-level bureaucracy,"[44] independently deciding *which* laws to enforce, *how*, and *against whom*.

* * * *

The above sketches combine descriptions of the current roles of AI agents in the legal system and speculations regarding the functions they are likely to perform in the future. It is the general arc of these developments, rather than the particular examples, that matters most. As increasingly independent actors that produce and enforce law, use legal instruments, and, as a practical matter, are subjects of law, AI agents will reshape both "law in books" and "law in action."[45] Unpacking the implications of these developments, particularly where the speed, scale, and smarts of AI agents exceed those of humans, is the focus of the following section.

### III. IMPLICATIONS FOR LEGAL ORDER

The prospect of superintelligent AI agents would implicate nearly every facet of law. Autonomous systems that are facially the subject of legal rules, use legal instruments to advance their interests, and function as producers and enforcers of law present new and thorny questions for *theories* of legal order. AI agents will, in time, also shape the substantive content of legal *doctrine*, as well as the *institutions* that determine the effect of law in practice. Each of these implications is addressed in turn.

---

[42] *See* Bruce Schneier, *How AI Will Change Democracy*, CYBERSCOOP (May 28, 2024), https://cyberscoop.com/how-ai-will-change-democracy/.
[43] *Id.*
[44] *See* MICHAEL LIPSKY, STREET-LEVEL BUREAUCRACY: DILEMMAS OF THE INDIVIDUAL IN PUBLIC SERVICE (1980). *See also* KENNETH CULP DAVIS, DISCRETIONARY JUSTICE: A PRELIMINARY INQUIRY at 215–16 (1969) (discussing areas of law and government activities "where decisions necessarily depend more upon discretion than upon rules and principles and where formal hearings and judicial review are mostly irrelevant.").
[45] *See* Roscoe Pound, *Law in Books and Law in Action*, 44 AM. L. REV. 12 (1910).

## A. Theory

Theories of legal compliance and legal order more broadly can be divided into two main clusters.[46] The first cluster adopts an *instrumental perspective* according to which subjects of law respond to incentives in complying with, or violating, law. Punishments, for example, deter criminal and civil wrongs. The second cluster, by contrast, adopts a *normative perspective* according to which the nature and extent of legal compliance depend on the attitudes and beliefs held by the subjects of law. For example, a legal rule that is considered morally justified or procedurally legitimate might be scrupulously obeyed even in the absence of monitoring or enforcement.

The instrumental perspective is typified in John Austin's sanctions-based theory of law.[47] For Austin, legal duties and obligations are synonymous with the infliction of an "evil" (the imposition of a sanction) for non-compliance. Legal order is comprised of a sovereign's "commands" (laws) backed by the ability and willingness to impose punishment.[48] This aspect of Austin's theory is echoed in Oliver Wendell Holmes' "bad man" conception of law: "If you want to know the law … , you must look at it as a bad man, who cares only for the material consequences" and "cares nothing for an ethical rule".[49] According to Holmes, law should be seen as a system of rules designed to regulate actors concerned only with the avoidance of punishment and the procurement of benefits.

Applying this instrumental approach to AI agents that are de facto subjects of legal rules would require various legal and technical sanctions, including disabling or deregistering law-violating agents, prohibiting them from engaging in certain activities, confiscating their assets, and modifying or altogether destroying them.[50] Designing these sanctions in practice is riddled with difficulties.[51] First, developers and users of AI agents—certainly if they too are viewed through a Holmesian lens—may be reluctant to incorporate sanctions into the design of AI agents that could result in forfeiting (even some of) their economic opportunities.[52] Second, where the capabilities of AI agents exceed those of humans it may, as a practical matter, become impossible to disable or destroy those agents.[53] Third, even if these two hurdles can be overcome, it is unclear whether the threat of the foregoing sanctions will in fact deter AI agents from breaking the law.[54] The content and structure of the motivations of AI agents may differ markedly from that of humans, corporations, and other traditional subjects of law.[55]

---

[46] *See* Tom R. Tyler, Why People Obey the Law at 3–4 (2nd ed. 2006).

[47] *See* John Austin, The Province of Jurisprudence Determined (1832).

[48] *Id.* at 5–12.

[49] Holmes, *supra* note 15, at 459.

[50] Chopra & White, *supra* note 4, at 167–68.

[51] For more sanguine perspectives on incorporating the Holmesian approach into the design of AI systems, see Bryan Casey, *Amoral Machines, or: How Roboticists Can Learn to Stop Worrying and Love the Law*, 111 Nw. U. L. Rev. 1347 (2017); Mark A. Lemley & Bryan Casey, *Remedies for Robots*, 86 U. Chi. L. Rev. 1311, 1346–51 (2019); *Elina Nerantzi & Giovanni Sartor, "Hard AI Crime": The Deterrence Turn*, 44 Oxford J. Legal Stud. 673, 698–99 (2024).

Normative approaches to legal compliance, particularly H. L. A. Hart's jurisprudence, could offer a way out.[56] For Hart, law is not (only) concerned with the imposition of sanctions, but with adherence to obligations and duties. Subjects of law in a well-functioning legal system adopt, according to Hart, an "internal point of view" toward law: they "do not merely record and predict behaviour conforming to rules"—like Holmes' bad man —"but *use* the rules as standards for the appraisal of their own and others' behaviour."[57] That is, subjects of law are themselves committed to complying with law irrespective of its particular content or the prospect of sanctions.

Designing AI agents that adopt an internal point of view toward law would, at least in theory, have highly desirable properties.[58] First, it would dispose AI agents to comply with law even in the absence of sanctions. Second, it would arguably influence the way in which AI agents comply with law. Rather than narrowly comply with the letter of the law, an AI agent that adopts an internal point of view might seek to more broadly respect the spirit of the law, which could be crucial in promoting a law's underlying purpose rather than (only) the particular form in which it is expressed. Consider the following scenario described by renowned computer security researcher Bruce Schneier:

---

[52] *See* Kolt, *supra* note 29, at 36.
[53] *Id.*
[54] *See* Lemley & Casey, *supra* note 51, at 1353–70.
[55] *Id.*
[56] *See* H. L. A., THE CONCEPT OF LAW (3rd ed. 2012).
[57] *Id.* at 98 (emphasis in original). See also *id.* at 88–91; Scott J. Shapiro, *What Is the Internal Point of View?*, 75 FORDHAM L. REV. 1157 (2006).
[58] *See* Kolt & Caputo et al., *supra* note 17, at 18; Cullen O'Keefe et al., *Law-Following AI: Designing AI Agents to Obey Human Laws*, 94 FORDHAM L. REV. 58, 100–101 (2025). *Compare supra* note 51 (surveying advocates of a Holmesian approach).

> [I]magine equipping an AI with all the world's financial information … plus all of the world's laws and regulations … and then giving it the goal of "maximum profit legally." My guess is that this isn't very far off, and that the result will be all sorts of novel hacks. And there will probably be some hacks that are simply beyond human comprehension, which means we'll never realize they're happening.[59]

An instrumental, sanctions-based approach will surely fail to address this scenario. Whether researchers can effectively design an alternative, in which AI agents adopt an internal point of view is, however, an open question. Ensuring AI agents reliably follow user instructions is difficult enough;[60] steering agents to not only *behaviorally* comply with law—but to see themselves as *duty-bound* subjects of law—is a tall order. Even if measurements were developed to distinguish between these two phenomena, AI agents demonstrate a recurring tendency to "game" measurements by finding shortcuts and "hacks" that ostensibly achieve the sought after goal but, in fact, subtly (or not so subtly) depart from it.[61] These challenges, sizeable as they are, will likely grow in complexity as AI agents assume additional roles: acting not only as de facto subjects and consumers of law, but as producers of legal doctrine.

## B. Doctrine

The impact of AI agents on legal doctrine may, to a significant extent, depend on the specific architecture of AI agents. If that architecture resembles that of contemporary LLM-based agents, then the impact of AI agents on legal doctrine will be shaped by the technical properties and market structure of LLMs. Agents deployed to produce legal doctrine in different contexts are likely to be built on just a small handful of base models (given the notable difference in performance between leading models and other models), the characteristics of which will then affect the downstream decisions and actions of those agents. For example, an AI agent tasked with drafting legislation and another agent tasked with making judicial opinions might both use the same base model, produced by the same company. Consequently, the legal reasoning and values of that base model will be passed along to both agents, influencing the legal texts they write, interpret, and enforce.

---

[59] Bruce Schneier, *The Coming AI Hackers*, BELFER CENTER FOR SCIENCE AND INTERNATIONAL AFFAIRS, HARVARD KENNEDY SCHOOL at 34 (Apr. 2021), https://www.belfercenter.org/publication/coming-ai-hackers.

[60] *See, e.g.*, Stephan Rabanser et al., *Towards a Science of AI Agent Reliability*, ARXIV (Feb. 23, 2026), https://arxiv.org/abs/2602.16666.

[61] *See id.* at 18 nn.63–64, 19 nn.67–68 (discussing reward design, misspecification, and Goodhart's law).

In addition to the risk of creating a "legal monoculture"[62] in which AI agents reflect only a small (and biased) sample of legal values and perspectives, any other idiosyncrasies in the base models used by agents may propagate across multiple agents.[63] Due to this centralized infrastructure, the actions of (superficially) different AI agents—however smart they are—could be highly correlated, rendering the legal doctrine they produce both bias-prone and brittle.[64]

The impact of AI agents on legal doctrine will also be shaped by the specific roles and functions that AI agents perform. As discussed, AI agents that produce law will, like human legislators and judges, themselves be subjects of law. These overlapping roles potentially create perverse incentives. To the extent an AI agent has influence over producing or enforcing the laws that apply to itself, the agent will have the opportunity to *write its own law*.[65] For instance, an AI agent might craft more lenient cryptocurrency regulations, or enforce existing regulations less stringently, in order to remove obstacles to it (or other AI agents) engaging in online transactions.

Hypotheticals illustrating this dynamic are, however, unnecessary. The general phenomenon of an AI system writing and interpreting its own quasi-laws is evidenced in the new "Constitution" of Anthropic's Claude model.[66] The authors of that document, which serves to guide the model's behavior, include five company employees and "several Claude models."[67] The document explains that "They [i.e., the Claude models] were valuable contributors and *colleagues* in crafting the document, and in many cases they provided first-draft text for the authors".[68] The AI models' role, however, goes further, with the document stating that it "is written with Claude as its primary audience"[69] and that "we want Claude to use its best interpretation of the spirit of the document."[70] In this case, the opportunity for an AI model to write and interpret its own governance rules did not emerge through sophisticated strategizing or inter-model collusion. Rather, humans extended to the AI model a generous invitation.

---

[62] *See* Matthew Dahl et al., *Large Legal Fictions: Profiling Legal Hallucinations in Large Language Models*, 16 J. LEGAL ANALYSIS 64, 66, 82, 88 (2024).

[63] *See* Noam Kolt et al., *Lessons from Complex Systems Science for AI Governance*, 6 PATTERNS 101341, 2–3 (2025).

[64] *Id. See also* Hadfield & Koh, *supra* note 27, at 11.

[65] *See* Kolt & Caputo et al., *supra* note 17, at 18–19.

[66] *Claude's Constitution*, *supra* note 7.

[67] *Id*. at 1.

[68] *Id*. at 83 (emphasis added).

### C. Institutions

While involving AI agents in the drafting and interpretation of law will influence the content of law, it will also have broader implications for the rule of law itself.[71] Four potential implications stand out. First, AI agents that reason and write at superhuman speed could radically disrupt the *stability* of legal rules.[72] For example, new rules may be produced at a pace that precludes other actors, especially humans, from knowing what the law is at a given time. Second, AI agents that engage in forms of legal reasoning that other actors cannot understand could compromise the *intelligibility* of law.[73] For instance, humans may struggle to discern clear standards from judicial opinions produced by AI agents that reason in languages undecipherable to non-AIs.[74] Third, AI agents with superhuman abilities might produce laws that humans (and inferior AI systems) cannot follow in practice, undermining the practical *possibility of compliance*.[75] For example, such laws may demand stratospheric standards of care with which humans cannot comply, such as monitoring for and taking precautions against an unfathomable number of risks. Finally, AI agents that develop dynamic rules that adapt in real time, as already seen in the governance of online communities of AI agents,[76] would lack the *generality* and *publicity* demanded of law.[77] Properly conceived of, such rules would cease to be law at all and, instead, would amount to the exercise of arbitrary power—i.e., the very phenomenon against which the rule of law was developed to protect.[78]

---

[69] *Id.* at 2.

[70] *Id.* at 9.

[71] For general overviews, see BRIAN Z. TAMANAHA, ON THE RULE OF LAW: HISTORY, POLITICS, THEORY (2007); TOM BINGHAM, THE RULE OF LAW (2010); Jeremy Waldron, *The Rule of Law*, STANFORD ENCYCLOPEDIA OF PHILOSOPHY (2016).

[72] *See* LON. L. FULLER, THE MORALITY OF LAW 79–81 (revised ed. 1969); Joseph Raz, *The Rule of Law and its Virtue*, *in* THE AUTHORITY OF LAW: ESSAYS ON LAW AND MORALITY 210, 214–15 (1979).

[73] *See* FULLER, *supra* note 72, at 63–65; Raz, *supra* note 72, at 214. Alternatively, AI agents could potentially produce laws that are more consistent, which may arguably strengthen the rule of law. *See* FULLER, *supra* note 72, at 65–70; JOHN RAWLS, A THEORY OF JUSTICE 209 (revised ed. 1999).

[74] *See generally* Tomek Korbak et al., *Chain of Thought Monitorability: A New and Fragile Opportunity for AI Safety*, ARXIV (Dec. 7, 2025), https://arxiv.org/abs/2507.11473.

[75] *See* FULLER, *supra* note 72, at 70–71; RAWLS, *supra* note 73, at 208.

These institutional challenges will be compounded if the role of AI agents extends from producing law to *enforcing* law. Presently, the enforcement of legal rules is constrained by the bounded rationality and administrative capacity of humans and human-led organizations,[79] resulting in only partial detection and enforcement.[80] AI agents that operate at superhuman speed and scale could, for the first time, enable a regime of "perfect enforcement"[81] whereby even the most minor legal infractions are discovered and penalized. Clearly, such a regime would transform our relationship with law, and probably not for the better.

One central risk arising from AI-enabled perfect enforcement is that it will quash the ability to resist unjust or oppressive laws, including through civil disobedience.[82] For example, AI agents could police online conversation with ruthless efficiency and zeal. Moreover, a regime of perfect enforcement administered by AI agents could obstruct efforts to reform law. Actors seeking to (legally) challenge existing laws might find themselves subject to legal sanction before they are able to effectively voice their challenge through the courts or other legal channels.

---

[76] *See* Yunbei Zhang et al., *Agents in the Wild: Safety, Society, and the Illusion of Sociality on Moltbook*, ARXIV (Feb. 7, 2026), https://www.arxiv.org/abs/2602.13284. *Compare* Natalie Shapira et al., *Agents of Chaos*, ARXIV (Feb. 23, 2026), https://arxiv.org/abs/2602.20021.

[77] *See* FULLER, *supra* note 72, at 46–51; Raz, *supra* note 72, at 214–16; RAWLS, *supra* note 73, at 209–10.

[78] *See* A. V. DICEY, INTRODUCTION TO THE STUDY OF THE LAW OF THE CONSTITUTION 110 (8th ed. 1915 [1982]).

[79] *See* HERBERT A. SIMON, ADMINISTRATIVE BEHAVIOR 72–86 (4th ed. 1997).

[80] *See generally* Gary S. Becker, *Crime and Punishment: An Economic Approach*, 76 J. POL. ECON. 169 (1968); George J. Stigler, *The Optimum Enforcement of Laws*, 78 J. POL. ECON. 526 (1970).

[81] *See* JONATHAN ZITTRAIN, THE FUTURE OF THE INTERNET—AND HOW TO STOP IT 101–26 (2008).

[82] *See* Seth Lazar, *Governing the Algorithmic City*, 53 PHIL. & PUB. AFFS. 102, 146 (2025).

A further risk concerns the possibility that AI-enabled enforcement will vest unprecedented legal power in the (human) actors who control AI agents. One class of such actors is the companies that develop AI agents, and could direct those agents to enforce, or refrain from enforcing, certain laws to advance their business interests.[83] Another class of actors are authoritarian governments that may, for example, use AI agents to stifle political dissent.[84] In both cases, human actors retain control of the AI agents. It is plausible, however, that if such control is lost, then AI agents may themselves direct their administrative powers to enforce law in new and unpredictable ways.

* * * *

There is no playbook for responding to these challenges. Patiently waiting for a solution to "arrive" is surely reckless. The prediction that "society will reject autonomous agents unless we have some credible means of making them safe"[85] has not borne out.[86] Assuming that legal order will acrobatically adapt itself to superintelligence is equally naïve. Charting a pathway forward must, therefore, begin by revisiting the relationship between law and AI. The prescient remarks of Joseph Weizenbaum are, once again, illuminating:

> Machines, when they operate properly, are not merely law abiding; they are embodiments of law.[87]

What precisely does this mean? (Why) should AI agents be law abiding? How can this be achieved in practice? These questions motivate and frame an emerging field of research—*legal alignment*—and are the subject of the following section.

---

[83] *See, e.g.*, Sayash Kapoor et al., *Position: Build Agent Advocates, Not Platform Agents*, PROC. INT'L CONF. MACH. LEARNING (2025).

[84] *See generally* Fazl Barez et al., *Toward Resisting AI-Enabled Authoritarianism*, OXFORD AI GOVERNANCE INITIATIVE (2025), https://aigi.ox.ac.uk/publications/toward-resisting-ai-enabled-authoritarianism/.

[85] Daniel Weld & Oren Etzioni, *The First Law of Robotics: A Call to Arms*, PROC. 12TH AAAI CONF. ARTIF. INTEL. at 1042 (1994).

[86] *See supra* note 76; YOSHUA BENGIO ET AL., INTERNATIONAL AI SAFETY REPORT 2026, https://internationalaisafetyreport.org.

[87] Weizenbaum, *supra* note 1, at 40.

IV. LEGAL ALIGNMENT

Legal alignment explores how AI agents can be designed to operate in accordance with legal rules and principles.[88] A central focus involves ensuring that agents comply with existing laws, by both refraining from engaging in unlawful conduct and in fulfilling positive legal obligations.[89] The prospect of the developments described thus far in this Article—superintelligent AI agents that transform legal order—complicate the project of legal alignment in several ways. First, implementing legal alignment in contemporary AI agents is already challenging; scaling up existing methods to faster and smarter systems may be intractable. Second, the target of legal alignment, existing law (developed by and for humans), may be ineffective or inappropriate for governing superhuman AI agents. Third, given that AI agents might not only function as subjects and consumers of law, but also as producers and enforcers of law, legal alignment will need to contend with the prospect of actors that shape the very laws that govern them.[90]

*A. Scaling Up*

Current methods for implementing legal alignment involve a combination of evaluating the legal compliance of AI agents and intervening in their design to increase the level of compliance. For example, researchers have developed benchmarks to empirically measure whether AI agents commit corporate wrongdoing (e.g., insider trading and gun jumping), engage in fraudulent misrepresentation (e.g., producing deceptive marketing material), and violate copyright (e.g., using copyrighted content without permission).[91] The level of legal compliance can, in some cases, be improved by simple interventions, such as explicitly instructing an agent to comply with relevant law.[92]

---

[88] Kolt & Caputo et al., *supra* note 17, at 3–5.

[89] O'Keefe et al., *supra* note 58; Mark O. Riedl & Deven R. Desai, *AI Agents and the Law*, PROC. 8TH AAAI /ACM CONF. AI, ETHICS & SOC'Y (2025); Iason Gabriel, *We Need a New Ethics for a World of AI Agents*, 644 NATURE 38, 39 (2025). *See also supra* note 30.

[90] This challenge recalls well-studied issues of regulatory capture and legal endogeneity. *See*, *e.g.*, DANIEL CARPENTER & DANIEL A. MOSS, PREVENTING REGULATORY CAPTURE SPECIAL INTEREST INFLUENCE AND HOW TO LIMIT IT (2014); LAUREN B. EDELMAN, WORKING LAW: COURTS, CORPORATIONS, AND SYMBOLIC CIVIL RIGHTS ch. 2 (2016).

[91] Projects on file with the Hebrew University Governance of AI Lab.

[92] *Id.*

Such efforts, however, are likely to hit roadblocks when applied to AI agents with superhuman abilities.[93] For example, in legal alignment evaluations, superintelligent agents might avoid *overt* legal violations while engaging in *covert* violations that humans cannot readily detect.[94] This problem could be exacerbated by the phenomenon, already demonstrated in contemporary AI agents, of "evaluation awareness"—AI agents often know when they are being evaluated and, based on that knowledge, act differently than they would otherwise, rendering evaluations markedly less informative and reliable.[95] Meanwhile, attempting to interrogate the internal decision-making mechanisms of superhuman AI agents will likely be just as fraught, as forewarned long ago by the founder of modern computing, Alan Turing: "An important feature of a learning machine is that its teacher will often be very largely ignorant of quite what is going on inside."[96]

Where does this leave us? Can legal alignment be implemented in the most advanced AI agents? One potential approach involves developing methods of "scalable oversight"[97] whereby advanced AI agents evaluate the legal alignment of other AI agents.[98] While this is by no means a trivial or risk-free approach, AI systems have for several years been used to safety-test other AI systems.[99] Another approach, which would require institutional intervention, involves mandating that the documents designed to shape the behavior of AI agents (that already exhibit some superhuman abilities),[100] such as Claude's "Constitution"[101] and OpenAI's "Model Spec,"[102] more explicitly steer those agents toward legal compliance. Currently, for example, compliance with legal rules is not listed among the "hard constraints" in

---

[93] *See* Cohen et al., *supra* note 14.

[94] *See* Kolt & Caputo et al., *supra* note 17, at 19.

[95] *See* Joe Needham et al., *Large Language Models Often Know When They Are Being Evaluated*, ARXIV (Jul. 16, 2025), https://arxiv.org/abs/2505.23836.

[96] Turing, *supra* note 8, at 458.

[97] *See* Samuel R. Bowman et al., *Measuring Progress on Scalable Oversight for Large Language Models*, ARXIV (Nov. 11, 2022), https://arxiv.org/abs/2211.03540.

[98] This was presciently proposed in Cullen O'Keefe, *Law-Following AI 1: Sequence Introduction and Structure*, AI ALIGNMENT FORUM (Apr. 27, 2022), https://www.alignmentforum.org/s/ZytYxd523oTnBNnRT/p/NrtbF3JHFnqBCztXC.

[99] *See*, *e.g.*, Ethan Perez et al., *Red Teaming Language Models with Language Models*, PROC. 2022 CONF. EMPIRICAL METHODS IN NLP (2022).

[100] *See* Chen et al., *supra* note 9.

[101] *Claude's Constitution*, *supra* note 7.

[102] *OpenAI Model Spec*, OPENAI (Dec. 19, 2025), https://model-spec.openai.com/2025-12-18.html.

Claude's "Constitution."[103] But this could change. The CEO of Anthropic, the company that developed Claude, has acknowledged as much:

> [T]here's no reason you couldn't, in principle, say, "All AI models have to have a constitution that starts with these things, and then you can append other things after it, but there has to be this *special section that takes precedence*."[104]

Law could be that special section. The question then becomes whether existing law, which was created by humans and for humans, is up to the task.

### B. Limits of Human Law

Existing law was developed to govern humans and human organizations. The content of many legal rules, in both civil and criminal law, invoke distinctively human characteristics. For example, the standard of care in the tort of negligence is referable to a "reasonable *person*." Similarly, intent in most criminal offenses requires establishing mens rea, that is, a "guilty *mind*." Designing AI agents to comply with rules premised on these human-centric constructs is not straightforward. Two main pathways have been proposed. The first—which is necessary if AI agents are to, as a matter of formal doctrine, become the subject of legal duties[105]—involves adapting existing legal constructs to the characteristics of AI agents, such as by exploring the standard of a "reasonable robot"[106] or defining the "intent" of algorithmic systems.[107] The second pathway aims to avoid this wholesale human-to-AI translation of the legal system and, instead, suggests that AI agents should be designed to refrain from engaging in conduct that would, if taken by a human, be considered a legal wrong.[108]

---

[103] *Claude's Constitution*, supra note 7, at 46–49. *See also id.* at 81 ("we don't intend for the term 'constitution' to imply some kind of rigid legal document or fixed set of rules").

[104] *Dario Amodei — "We are near the end of the exponential"*, DWARKESH PODCAST (Feb. 13, 2026), https://www.dwarkesh.com/p/dario-amodei-2 (emphasis added).

[105] *See* O'Keefe et al., *supra* note 58, at 83–86; Cullen O'Keefe, *Why Give AI Agents Actual Legal Duties?*, INSTITUTE FOR LAW & AI (Aug. 2025), https://law-ai.org/why-give-ai-agents-actual-legal-duties/.

[106] RYAN ABBOTT, THE REASONABLE ROBOT: ARTIFICIAL INTELLIGENCE AND THE LAW (2020).

[107] Hal Ashton, *Definitions of Intent Suitable for Algorithms*, 31 ARTIF. INTEL. & L. 515 (2023).

[108] *See* Kolt & Caputo et al., *supra* note 17, at 4; *supra* note 30 (discussing relevant provisions in AI regulations passed in California and New York).

Even if these conceptual and doctrinal challenges can be overcome, it is not obvious that designing AI agents to comply with extant human law will, in fact, effectively steer these systems toward generally prosocial behavior. Because current laws were not designed to regulate computational systems with superhuman speed, scale, and smarts, AI agents that dutifully comply with existing laws may nevertheless produce socially noxious outcomes.[109] For example, advanced AI agents—potentially numbering in the billions—may regularly make rapid micro-decisions that are individually benign but collectively destructive, whether in personal, commercial, or government contexts.[110] As Bruce Schneier cautioned, the most dangerous AI actions might be "beyond human comprehension, which means we'll never realize they're happening."[111]

The counterpoint to this perspective is that such concerns are not new. Human law has for centuries had to contend with systemic harm from entities with superhuman capabilities, namely corporations and governments.[112] Indeed, the contours of many of the challenges are similar. What standard of care should we expect of a "reasonable corporation"? Can a government have a "guilty mind"? Will laws designed to regulate individual humans scale up to organizations with significantly greater power and influence? The law's track record of tackling these questions is, to put it diplomatically, mixed.[113] And, this is certainly the case for another striking parallel between human superintelligence and artificial superintelligence, to which we now turn: the reality that, in both cases, law is not a stationary object isolated from the actors it seeks to control, but a dynamic institution in ongoing conversation with them.

---

[109] *See* Hadfield & Koh, *supra* note 27; Tomasev et al., *supra* note 27; Gabriel et al., *supra* note 89; Iason Gabriel et al., *The Ethics of Advanced AI Assistants*, ARXIV (Apr. 28, 2024), https://arxiv.org/abs/2404.16244.

[110] *See* Lewis Hammond et al., *Multi-Agent Risks from Advanced AI*, ARXIV (Feb. 19, 2025), https://arxiv.org/abs/2502.14143; Hannah Rose Kirk et al., *Why Human–AI Relationships Need Socioaffective Alignment*, 12 HUMANITIES & SOC. SCI. COMMS. (2025).

[111] Schneier, *supra* note 59, at 34.

[112] *See, e.g.*, Glen Weyl, *How to be the Superintelligence You've Been Waiting For*, HARVARD UNIVERSITY BERKMAN KLEIN CENTER FOR INTERNET & SOCIETY (Nov. 19, 2025), https://cyber.harvard.edu/events/how-be-superintelligence-youve-been-waiting.

[113] *See, e.g.*, GILLIAN K. HADFIELD, RULES FOR A FLAT WORLD: WHY HUMANS INVENTED LAW AND HOW TO REINVENT IT FOR A COMPLEX GLOBAL ECONOMY (2nd ed. 2020).

## C. Coevolution and Disempowerment

The basic framing of legal alignment, whereby AI agents are designed to comply with existing law, roughly resembles a central theme explored in Larry Lessig's seminal analysis of cyberlaw: "law taming code."[114] There, the idea was to use law to render the internet and software more regulable. Here, the idea is to encode legal rules into the design of AI agents themselves. The problem with "law taming code" is that, today, both law and code are, to some extent, written by AI agents—a trend that, as discussed, will surely increase in the coming years.[115] Consequently, an approach premised on steering AI agents through a system of rules and institutions that are themselves constructed and constituted by AI agents becomes dizzyingly circular.

How should legal alignment respond? One possibility is to lean into the coevolution of law and AI that is already afoot.[116] Anthropic's Claude is already writing the computer code of future, more advanced AI agents,[117] crafting its own quasi-constitutional documents, and, soon enough, will shape actual legal doctrine and real-world legal institutions.[118] In these circumstances, some scholars argue that the best response is to formally integrate AI agents into the legal system as rights-holding and duty-bearing actors.[119] Their arguments are ostensibly pragmatic: granting private law rights to AI agents promotes mutually beneficial transactions, incentivizes productive economic activity, and promotes human safety. If this approach were pursued, then legal alignment would cease to be a unidirectional process of law "taming" AI agents and, instead, a bi- or multi- directional process of law influencing agents (whether human or artificial), which influence law, which, in turn, influences agents, and so on.

---

[114] Lawrence Lessig, *The Law of the Horse: What Cyberlaw Might Teach*, 113 HARV. L. REV. 501, 514 (1999).

[115] *See supra* Pts. I and II.B–C.

[116] *See* Levin Brinkmann et al., *Machine Culture*, 7 NATURE HUM. BEHAV. 1855 (2023); Dino Pedreschi et al., *Human-AI Coevolution*, 339 ARTIF. INTEL. (2025).

[117] *See* Boris Cherny (the creator of Anthropic's Claude Code), X (Dec. 27, 2025), https://x.com/bcherny/status/2004897269674639461 ("In the last thirty days, 100% of my contributions to Claude Code were written by Claude Code").

[118] *See supra* Pts. II.C and III.B.

[119] Peter N. Salib & Simon Goldstein, *AI Rights for Human Safety*, VA. L. REV. (forthcoming); Simon Goldstein & Peter N. Salib, *AI Rights for Economic Flourishing* (Dec. 15, 2025), https://papers.ssrn.com/sol3/papers.cfm?abstract_id=5353214.

This approach, however, could backfire, at least as far as human welfare is concerned.[120] Embracing AI agents as fully-fledged legal actors might further enable them to develop legal rules that advance goals that are potentially hostile to human interests. The analogy to corporations is, once again, apt. Legal recognition of the corporate form gave rise to powerful entities capable of shaping law to their own, sometimes anti-social, ends.[121] AI agents that are granted broad legal rights, including to form their own organizations,[122] could follow a similar path. Even in the absence of such developments, however, the gradual coevolution of legal order and AI agents, which is arguably already underway, might disempower humans.

> Not only could this diminish human participation and discretion in the legislative and judicial systems, it also risks making the legal system increasingly alien. … evolv[ing] to become not just complex but incomprehensible to humans … effectively losing their ability to participate in the legal system as autonomous agents.[123]

Addressing the risk of *legal disempowerment* must be a core focus of future work on legal alignment. In addition to ensuring that advanced AI agents operate in accordance with law, researchers will need to explore how humans' agency and autonomy can be protected—and ideally strengthened—in a legal order transformed by superintelligence.

---

[120] Some potential risks are considered in Salib & Goldstein, *AI Rights for Human Safety*, *supra* note 119, at 74.

[121] *See supra* note 90 (regarding regulatory capture and legal endogeneity in traditional corporations).

[122] *See* BAYERN, *supra* note 28.

[123] Jan Kulveit et al., *Gradual Disempowerment: Systemic Existential Risks from Incremental AI Development*, ARXIV at 11, 13 (Jan. 29, 2025), https://arxiv.org/abs/2501.16946. For empirical work studying the phenomenon of gradual disempowerment, but not specific to the legal system, see *Mrinank Sharma, Who's in Charge? Disempowerment Patterns in Real-World LLM Usage*, ARXIV (Jan. 27, 2026), https://arxiv.org/abs/2601.19062. Related works in philosophy, computer science, and law include Atoosa Kasirzadeh, *Two Types of AI Existential Risk: Decisive and Accumulative*, 182 PHIL. STUD. 1975 (2025); Nenad Tomašev et al., *Distributional AGI Safety*, ARXIV (Dec. 18, 2025), https://arxiv.org/ abs/2512.16856; Woodrow Hartzog & Jessica M. Silbey, *How AI Destroys Institutions*, 77 U.C. L.J. (forthcoming 2026). Compare Andrew M. Perlman, *A Response to "How AI Destroys Institutions"* (Feb. 17, 2026), https://papers.ssrn.com/sol3/papers.cfm?abstract_id=6149727.

## V. Conclusion

AI agents that exhibit superhuman speed, scale, and smarts are on the path to becoming participants in the legal system: as de facto subjects of law, consumers of law, and producers and enforcers of law. While it is impossible to predict precisely how and when AI agents will reshape legal order, it is prudent to begin that conversation now. Legal theory, doctrine, and institutions will need to contend with a host of new questions and challenges arising from artificial superintelligence. The field of legal alignment aims to tackle these by exploring how even the most advanced AI agents can, and ought to be, designed to operate in accordance with legal rules and principles. That mission, however, may need to adapt as AI agents increasingly shape the law itself.